\documentclass[aip,jcp,preprint]{revtex4-1}   
 
\usepackage{graphicx}%
\usepackage{dcolumn}%
\usepackage{bm}%
\usepackage{color}
\usepackage{graphics,epsfig}
\usepackage{amsmath}
\usepackage{multirow}                              
\usepackage{setspace} 	                         
\usepackage{epsfig}
\usepackage{xspace}
\usepackage[export]{adjustbox}
\usepackage{epstopdf}

\input colordvi.tex
\graphicspath{{./}{figures/}}

\begin{document}

\setcounter{page}{1}

\title{\vspace{-15mm}\fontsize{24pt}{10pt}\selectfont\textbf{Article Title}} 
\title{Coulomb crystal mass spectrometry in a digital ion trap} 
\author{Nabanita Deb*}
\affiliation{Department of Chemistry, University of Oxford, Chemistry Research Laboratory, 12 Mansfield Road, Oxford, OX1 3TA, United Kingdom}
\author{Laura L. Pollum*}
\affiliation{Department of Chemistry, University of Oxford, Chemistry Research Laboratory, 12 Mansfield Road, Oxford, OX1 3TA, United Kingdom}
\author{Alexander D. Smith}   
\affiliation{Department of Chemistry, University of Oxford, Chemistry Research Laboratory, 12 Mansfield Road, Oxford, OX1 3TA, United Kingdom}
\author{Matthias Keller}
\affiliation{Department of Physics and Astronomy, University of Sussex, Brighton, BN1 9QH, United Kingdom}
\author{Christopher J. Rennick}
\affiliation{Department of Chemistry, University of Oxford, Chemistry Research Laboratory, 12 Mansfield Road, Oxford, OX1 3TA, United Kingdom}
\author{Brianna R. Heazlewood}
\affiliation{Department of Chemistry, University of Oxford, Chemistry Research Laboratory, 12 Mansfield Road, Oxford, OX1 3TA, United Kingdom}
\author{Timothy P. Softley}
\affiliation{Department of Chemistry, University of Oxford, Chemistry Research Laboratory, 12 Mansfield Road, Oxford, OX1 3TA, United Kingdom}
\email{tim.softley@chem.ox.ac.uk}  

\date{\today}

\begin{abstract}
We present a mass spectrometric technique for identifying the masses and relative abundances of Coulomb-crystallized ions held in a linear Paul trap. A digital radiofrequency waveform is employed to generate the trapping potential, as this can be cleanly switched off, and static dipolar fields subsequently applied to the trap electrodes for ion ejection. Excellent detection efficiency is demonstrated for {Ca}$^+$ and {CaF}$^+$ ions from bi-component {Ca}$^+$/{CaF}$^+$ Coulomb crystals 
 prepared by reaction of Ca$^+$ with CH$_3$F. 
A quantitative linear relationship is observed between ion number and the corresponding integrated TOF peak, independent of the ionic species. The technique is applicable to a diverse range of multi-component Coulomb crystals 
 - demonstrated here for Ca$^+$/NH$_3^+$/NH$_4^+$ and Ca$^+$/CaOH$^+$/CaOD$^+$ crystals -
and  will facilitate the measurement of ion-molecule reaction rates and branching ratios in complicated reaction systems.
\end{abstract}


\maketitle

\section{Introduction}
Recent improvements in cold molecular beam methods and their combination with ion traps is facilitating exciting new developments in the observation of cold, controlled ion-molecule reactions.\cite{ARPC,Willitsch_intrevPC}  It is not sufficient, however, to simply conduct experiments in the cold regime; the resulting products must be quantitatively detected to extract information about the reaction process, such as the rate, product branching ratio, and energy dependence.  

\begin{figure}[htb]
\centering
\vbox{
    \includegraphics[trim = 0mm 160mm 40mm 0mm, clip,width=1.0\textwidth]{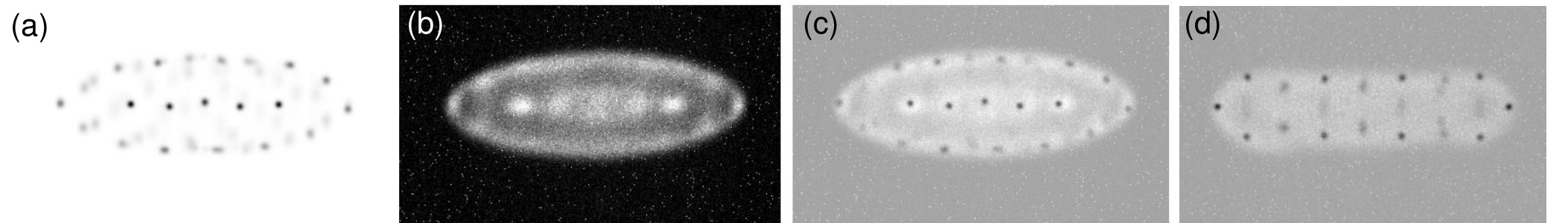}
}
\caption{The number of ions in a given Ca$^+$ Coulomb crystal can be established to within $\pm 5$ ions by fitting the experimental image to MD simulations. A simulated crystal image with 50 Ca$^+$ ions is shown in panel (a), alongside an experimental Ca$^+$ crystal image in panel (b). The simulated crystal is overlaid on the experimental image in (c), illustrating the excellent agreement between the positions of the ions in the simulated and experimental crystals. This approach is also adopted for multi-component crystals; in panel (d), a bi-component Ca$^+$/CaF$^+$ experimental crystal (in which only the Ca$^+$ are fluorescing) is overlaid with a simulated Ca$^+$/CaF$^+$ crystal comprising 30 Ca$^+$ and 35 CaF$^+$ ions. Only the Ca$^+$ ions are made visible in the simulation, for ease of comparison between the images.}
\label{fig:CCMD_sims}
\end{figure}

Trapped, laser-cooled ions can undergo a phase transition to adopt a highly ordered `Coulomb crystal' structure. Co-trapped ions of a different species may be sympathetically cooled through elastic collisions with the laser-cooled ions, forming multi-component Coulomb crystals.  In this way, cold molecular ion targets of species that cannot be laser cooled can be prepared. Imaging the fluorescence continuously emitted by the laser-cooled ions enables real-time observation of the crystal framework, as depicted in figure \ref{fig:CCMD_sims}(b).
Due to the dependence of the trapping field on the charge-to-mass ratio of the ions, different ionic species are located in separate regions of the Coulomb crystal. In chemical reactions, this separation can be exploited and reaction rates measured through monitoring the changing shape of the crystal. For instance, the reaction of laser-cooled ions with neutral reactants -- and the disappearance of fluorescing ions --
is accompanied by changes in the crystal framework as product ions heavier (lighter) than the laser-cooled species migrate to the outer (inner) region of the crystal.\cite{ARPC,WillitschPCCP} 
Hence, even though the non-fluorescing sympathetically cooled ions can not be observed directly,
  comparison of molecular dynamics (MD) simulations with observed fluorescence images enables quantitative determination of ion numbers
  (see Fig. \ref{fig:CCMD_sims}).

In complicated chemical systems with multiple product channels, however, MD simulations alone are insufficient for establishing accurate reaction rates.
For example, in the reaction between C$_2$H$_2^+$ (sympathetically cooled into a {Ca}$^+$ Coulomb crystal) and NH$_3$, there are numerous exothermic pathways: proton transfer, charge transfer, two H-elimination pathways and two H$_2$-elimination pathways.\cite{Anderson} 
In this case, both the reactant ions and numerous `dark' (non-fluorescing) product ions are lighter than the laser-cooled species, and thus would be indistinguishable in the dark inner core.  
A change in the ratio of the numbers of light ions of different mass would have little effect on the crystal framework.
In some instances, resonant-excitation mass spectrometry (MS) methods can be utilized.  
The scanning of an additional radiofrequency (RF) field resonantly excites the secular motional frequencies of trapped ions, allowing the ion masses to be determined.\cite{Baba,Welling,Mortensen}  In the limiting case of a single molecular ion sympathetically cooled by a single laser-cooled ion, the mass of the molecular ion can be precisely measured, with for example $^{26}$MgH$^+$ and $^{26}$MgD$^+$ experimentally distinguishable.\cite{Staanum}  While several variants of this non-destructive technique have been developed,\cite{Roth,Raab} resonant-excitation MS remains limited in application to small crystals with only a few different ion types; when motional frequency spectra are recorded for large multi-component crystals, the features are 
poorly resolved, difficult to unambiguously assign, and the relative abundances cannot be ascertained quantitatively.\cite{WillitschPCCP}  
An alternative approach, termed ion crystal weighing, employs a pulsed electric field to excite the center-of-mass mode of the ions.\cite{Sheridan}  The frequency of the mode, and hence the mass of the crystal, can be deduced from the Fourier-transformed autocorrelation of the ion fluorescence signal.  
While ion crystal weighing is applicable to larger crystals, interpretation of the results is challenging when 
more than 2 ionic species are present or when the identities of the dark ions are unknown.

Adopting the principles of Wiley-McLaren time-of-flight MS (TOF-MS),\cite{WileyMcLaren} Hudson and co-workers recently demonstrated the ejection of ions from a linear Paul trap into a TOF tube and onto an external detector.\cite{Hudson}  Ion ejection was achieved by switching off the RF voltages and converting the trap electrodes into a repeller-extractor pair by applying static voltages. 
However,
when using resonantly-driven cosine RF voltages, the trap electrodes form part of a resonant inductor-capacitor circuit.  As the voltage is switched from cosine RF to direct current (DC), the amplitude of the trapping voltages decays with a time constant $1/k$ on the order of microseconds, and ringing of the form $V_{\text{rf}}$e$^{-kt}$cos$(\omega t)$ is superimposed on the ejection voltages. This compromises detection efficiency; ions experience vertical deflection away from the TOF axis and typically miss the detector.  In an attempt to overcome this problem, Hudson and co-workers applied very high repeller and extractor voltages (1.4~kV and 1.2~kV, respectively).
The 
observed 
mass-spectrum peak intensities, however, were not representative of the relative quantities of each species.\cite{Hudson}

Quadrupole ion traps have been successfully combined with mass analysis apparatus in a variety of configurations -- with ion ejection both along and perpendicular to the ion trap axis -- and many such instruments are now commercially available.\cite{Lubman,Orbitrap} 
Such systems, however, are not designed for the study of ion-molecule reactions in Coulomb crystals.  There is insufficient optical access to the trap for effective laser cooling and crystal imaging.  Furthermore, the typical operating pressures and repeated ejection cycles are not appropriate for the long storage times required in our experiments -- we require ultra-high vacuum conditions to achieve ion storage times on the order of tens of minutes.

In this paper, we present a new digital ion trap (DIT) Wiley-McLaren MS technique for identifying the masses and relative abundances of all ions within a Coulomb crystal.  
Digital trapping waveforms, also termed square-wave or pulsed waveforms, are applied to a linear Paul trap in place of the conventional cosine waveform. This facilitates rapid, clean switching between the RF trapping and static DC ejection voltages, with no residual ringing.  The entire crystal is ejected radially from the trap, where the ions pass through a TOF tube onto a microchannel plate (MCP) detector for mass-sensitive detection.  The quantitative performance of the technique is characterized by the 
analysis of hundreds of {Ca}$^+$ and {Ca}$^+$/{CaF}$^+$ Coulomb crystals, in addition to multi-component {Ca}$^+$/{CaOH}$^+$/{CaOD}$^+$ and {NH}$_3^+$/{NH}$_4^+$/{Ca}$^+$ crystals.  The efficiency of ion ejection is studied, aided by extensive modeling. 
The methodology introduced in this work is applicable to any reaction of laser-cooled or sympathetically-cooled ions, provided two conditions are satisfied: the co-trapped ions 
have an appropriate mass-to-charge ratio for efficient sympathetic cooling; and any product ions 
are formed with insufficient kinetic energy to escape the trap.

\section{Experimental Methods}

\begin{figure}[htb]
\centering
\vbox{
    \includegraphics[trim = 0mm 100mm 80mm 5mm, clip,width=0.5\textwidth]{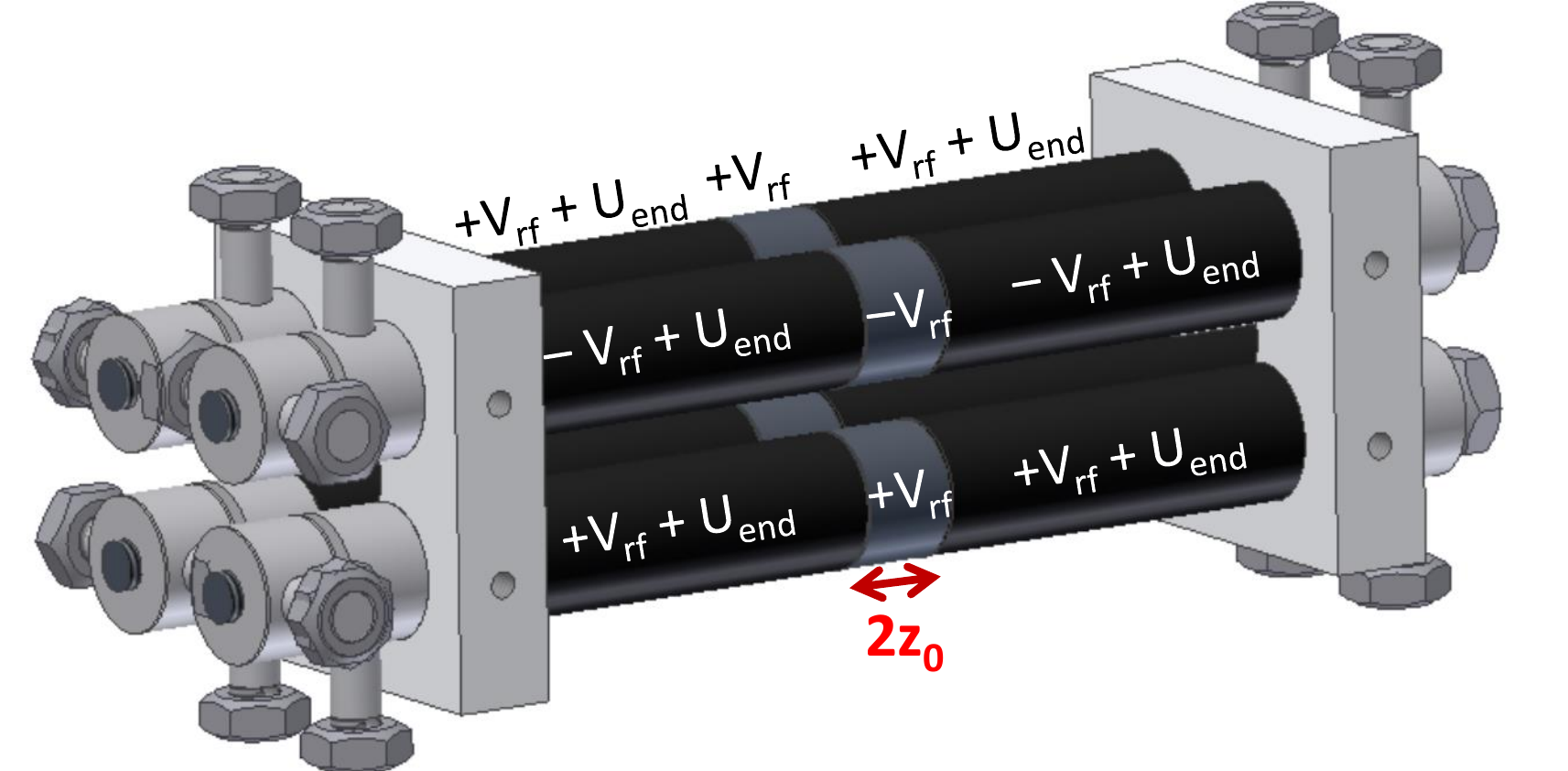}
}
\caption{(color online). Schematic illustration of the linear Paul trap utilized in this work, with four electrodes in a quadrupole arrangement.  Each electrode is composed of three segments, with RF voltages ($V_{\text{rf}}$) applied to all segments and additional DC voltages ($U_{\text{dc}}$) applied to the endpieces.}
\label{fig:ion_trap}
\end{figure}

The experiments presented in this paper utilize a linear Paul trap operated in ultrahigh vacuum (UHV) conditions, illustrated in Fig. \ref{fig:ion_trap}.
The cylindrical  trap electrodes, radius 4~mm, have a diagonal electrode-surface separation of 
$2r_0 = 7.0$~mm   and are each composed of three segments (endcap separation 2$z_0= 5.5$~mm), to which a combination of radiofrequency (RF) and static voltages are applied.
The conventional time-dependent trapping potential in a linear Paul trap is of the form $\phi_{\text{rf}}(x,y,t) = \frac{V_{\text{rf}}}{2} \left(\frac{x^2 - y^2}{r_0^2}\right) {\rm cos} (\omega_{\text{rf}} t)$,
with $V_{\text{rf}}$ the peak-to-peak voltage amplitude, $\omega_{\text{rf}}$ the RF drive frequency.
However, as first demonstrated in the 1970s,\cite{Richards,Sheretov} a cosine wavefunction is not essential for the generation of a trapping potential.  A DIT, as we utilize in this work, employs a rectangular waveform of period $T$ and fractional pulse width $\tau$, yielding a time-dependent trapping potential\cite{Drewsen_digital1,Drewsen_digital2} 
\begin{equation}
\phi_{\text{rf}}(x,y,t) = \frac{V_{\text{rf}}}{2} \left(\frac{x^2 - y^2}{r_0^2}\right) P_{\tau} (t),
\end{equation}
\begin{equation}
P_{\tau} (t) = \left\{
\begin{array}{lr}
1 & \text{ if } \; |t| \le \tau T/2 \\
0 & \text{ if } \; \tau T/2 < |t| \le (1-\tau)T/2 \\
-1 & \text{ if } \; (1-\tau)T/2 < |t| \le (1+\tau)T/2 \\
0 & \text{ if } \; (1+\tau) T/2 < |t| \le (2-\tau)T/2 \\
1 & \text{ if } \; (2-\tau) T/2 < |t| \le T \\
\end{array}
\right.
\end{equation}
\begin{equation}
P_{\tau}(t+T) = P_{\tau} (t),
\end{equation}
where we can represent the pulsed waveform as the sum of its cosine Fourier components, $P_{\tau} (t) = \sum\limits_{n=1}^{\infty} a_n {\rm cos} (n\omega_{\text{rf}} t)$.

Axial confinement
of ions in the trap is obtained through the application of static voltages ($U_{\text{dc}}$) to the endpieces 
of the segmented electrodes, 
$\phi_{\text{end}}(x,y,z) = \frac{\eta U_{\text{dc}}}{z_0^2} \left(z^2 - \frac{x^2 + y^2}{2}\right)$,
with $\eta=0.244$ a geometrical factor for this trap. The total electric potential is thus 
$\phi = \phi_{\text{rf}}(x,y,t) + \phi_{\text{end}}(x,y,z)$.
The equations of motion for a single trapped ion can be expressed as a Mathieu equation, ${d^2 u}/{d \xi ^2} + u[a_u + 2q_u f(\xi)] = 0$, where $f(\xi)={\rm cos}(2\xi)$ for a cosine RF field and $f(\xi)= P_{\tau} (\xi)$ for a digital RF field,  $u \in [x,y]$ and $\xi = \frac{1}{2} \omega_{\text{rf}} t$.  The stability of an ion trajectory in the trap is dependent on the dimensionless Mathieu parameters $a_u$ and $q_u$. The $(q_u,a_u)$ plane is divided into stable and unstable regions; for stable values of $a_u$ and $q_u$, the solutions to the Mathieu equations are bound and the ion will be confined in the trap. In a DIT, the Mathieu equations are dependent on the fractional pulse width ($\tau$), which can be varied to yield a stability diagram comparable to that of a cosine trap.  (See reference~[\onlinecite{Drewsen_digital1}] for a derivation of the stability criterion in a DIT.) Bandelow \emph{et al.} have recently experimentally confirmed the applicability of the $a_u$ and $q_u$ parameters in mapping the stability regions of a hyperbolic Paul DIT.\cite{Bandelow}

\begin{figure}[htb]
\centering
\includegraphics{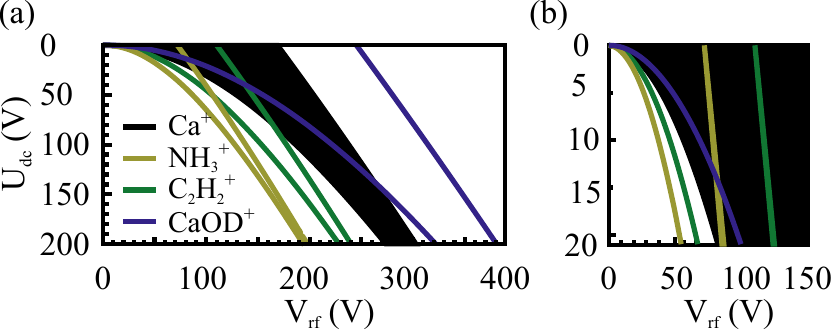}
\caption{(color online). (a) The regions of stability for several ionic species in the DIT operating at $\tau$ = 0.25 and $\omega_{\text{rf}} = 2 \pi \times 1.33$~MHz. For clarity, only the stability region of Ca$^+$ is shaded black; for all other species, the stability region is enclosed between the relevant colored (shaded) curves.  Regions of overlapping stability indicate the operating parameters ($V_{\text{rf}}, U_{\text{dc}}$) where species can be co-trapped.  (b) A closer view of the experimentally-accessible region of the stability diagram.}
\label{fig:stability_diagram}
\end{figure}

We can define dimensionless Mathieu parameters which are dependent on the mass ($m$) and charge ($Q$) of the species,\cite{Willitsch_intrevPC,Drewsen_digital1}
\begin{equation}
a_x = a_y = -\frac{a_z}{2} = - \frac{\eta 4 Q U_{\text{dc}}}{m z_0^2 \omega_{\text{rf}}^2},
\end{equation}
\begin{equation}     
q_x = -q_y = - \frac{2 Q V_{\text{rf}}}{m r_0^2 \omega_{\text{rf}}^2}; \,q_z = 0,
\end{equation}
which enable one to construct a DIT stability diagram in terms of $U_{\text{dc}}$ and $V_{\text{rf}}$ (see Fig. \ref{fig:stability_diagram}).
From the secular potential, defined as the sum of the pseudo-potential (the harmonic time-averaged potential arising from the oscillating RF fields) and the static endpiece potential, one can derive the digital trap depth,
\begin{equation}
\phi_{\text{sec}}(r_0) = c \frac{Q^2}{2 m \omega_{\text{rf}}^2} \frac{V_{\text{rf}}^2}{r_0^2} - \frac{\eta Q U_{\text{dc}} r_0^2}{2z_0^2},
\end{equation}
with $c = \frac{1}{2} \sum\limits_{n=1}^{\infty} a_n^2 / n^2$ a sum over the Fourier-component amplitudes $a_n$ (in a cosine trap, $c = 0.5$).
The digital RF voltage is applied at a frequency of $\omega_{\text{rf}} = 2 \pi \times 1.33$~MHz, with typical trap parameters $V_{\text{rf}} = 50$~V, $U_{\text{dc}} = 1.0$~V and $\tau = 0.25$ that 
give Matthieu parameters falling
within the regions of stability for all ionic species considered in this work. 

\begin{figure}[htb]
\centering
\includegraphics{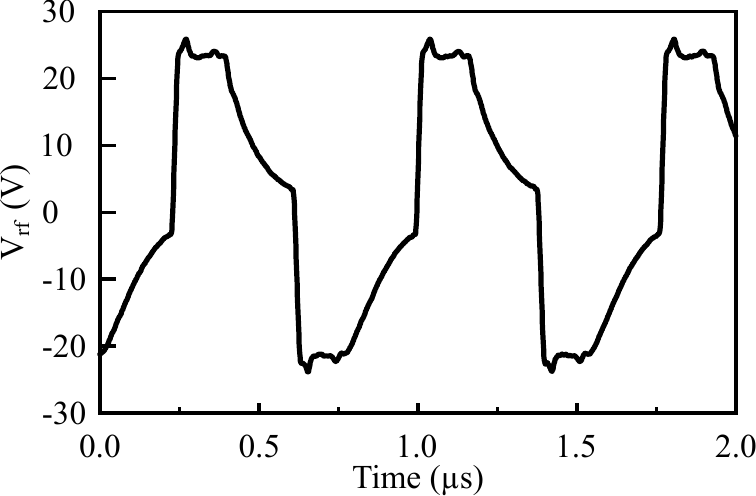}
\caption{Experimental digital RF trapping waveform.}
\label{fig:digitalwaveform}
\end{figure}

The experimental digital waveform (Fig. \ref{fig:digitalwaveform}) exhibits a non-zero pulse decay time, and thus deviates from the ideal digital waveform. MD simulations -- in which the measured digital waveform can be directly utilized -- indicate that this deviation from the ideal waveform has a very minor impact on the trap depth and  does not adversely affect crystal formation, crystal stability or the ejection properties of the system. This is in agreement with Sudakov \emph{et al.}, who found non-ideal waveforms to have an insignificant effect on the calculated regions of stability in a DIT.\cite{Sudakov}
Hence assuming the real digital waveform is well represented by an ideal digital waveform, a trap depth of $\sim$1.21~eV is calculated -- shallower than the cosine trap depth of $\sim$1.36~eV as previously utilized in our research
 group \cite{Willitsch_PRL},
but entirely sufficient for most applications.

To form Coulomb crystals, calcium atoms in an effusive skimmed beam are non-resonantly ionized (at 355~nm) in the trap center.  The resulting {Ca}$^+$ ions are laser cooled on the $4s$~$^2S_{1/2} \rightarrow 4p$~$^2P_{1/2}$ transition at 397~nm, with a 866~nm repump beam addressing population lost to the $3d$~$^2D_{3/2}$ state. The cooling lasers are reflected back along the trap axis to achieve bi-directional cooling.   An additional 397~nm beam at 45$^{\circ}$ to the trap axis provides radial cooling.  Fluorescence continuously emitted by laser-cooled {Ca}$^+$ ions is imaged by a lens into an intensified charge-coupled device camera system, yielding a two-dimensional image of the Coulomb crystal.  High-precision leak valves facilitate the controlled introduction of neutral reactants into the ion trap chamber, typically held at pressures $\sim 1 \times 10^{-9}$~mbar. In this way, we generate molecular ions through chemical reactions between the neutral reactant gas and trapped Ca$^+$ ions,
 or with trapped sympathetically cooled ions.

\begin{figure}[htb]
\centering
\vbox{
    \includegraphics{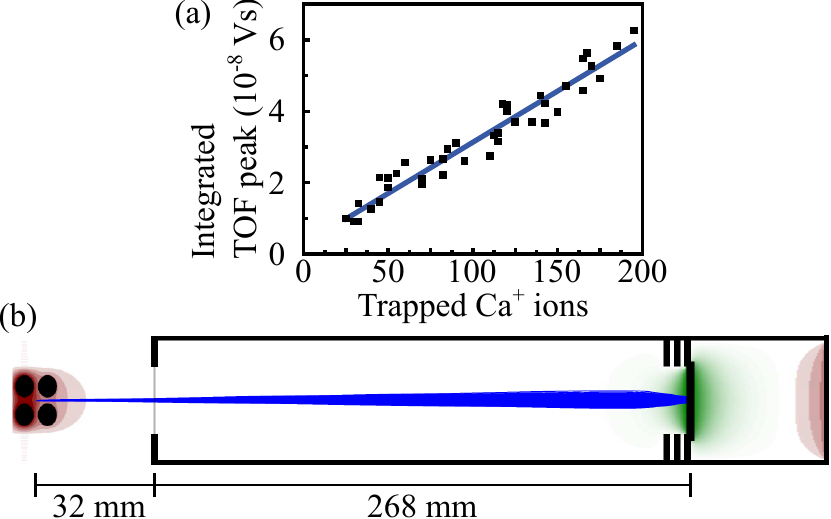}
}
\caption{(color online). (a) The integrated Ca$^+$ peak intensity in the TOF spectrum is plotted for each Coulomb crystal against the  number of Ca$^+$ ions, determined from fitting the experimental images to MD simulations.  (b) Schematic illustration of the TOF-MS apparatus, depicting the ion trap electrodes (viewed along the trap axis, with ion ejection perpendicular to this axis), TOF tube and MCPs. The electric potential 
near the ion trap is positive (shaded red online), and near the front detector plate is negative (shaded green online)
 with the trajectories of 200 Ca$^+$ ions ejected from the trap shown (in blue online.)}
\label{fig:realidealtrappingvoltage}
\end{figure}

After the quadrupole trapping voltages are switched off, ions are radially ejected into the TOF tube by dipolar repeller and extractor fields of +220~V and +97~V respectively, applied with a rise time of 70~ns,
 and are accelerated towards a grounded mesh of $88\%$ transparency mounted across the entrance of the TOF tube (see Fig. \ref{fig:realidealtrappingvoltage}(b)). The ions are  detected by a microchannel plate (MCP) detector 
at the end of the flight tube. 
The repeller and extractor voltages employed were found empirically to give the best TOF mass resolution for {Ca}$^+$,
and this two-stage acceleration arrangement is equivalent to a Wiley-McLaren time-of-flight scheme \cite{WileyMcLaren}. 
In the experiments reported here there was (nominally) zero delay in the switching from the trapping voltages to the ejection voltages; although the delay can be varied easily in the digital trap, minimizing the delay was found to optimize the mass resolution. 

\section{Simulations}
Coulomb crystals are simulated using a custom-written MD crystal simulation program, which takes the experimental digital waveform as the input RF trapping voltage.
Simulated crystal images can thus be directly compared to those observed experimentally, as demonstrated in Fig. \ref{fig:CCMD_sims}.  A MATLAB program is used to model the ion trajectories upon ejection by calculating the electric fields in the ion trap and TOF tube, using SIMION to solve the Laplace equation numerically and including ion-ion repulsion.  Taking the identity, position and velocity of each ion within a Coulomb crystal (as output by the MD code) and the experimental ejection voltages as input parameters, ion trajectories are calculated using a Velocity-Verlet algorithm. For Coulomb crystals comprised of $\le$ 200 ions, simulations indicate that the detection efficiency is limited only by the transmittance of the grounded mesh; detection efficiency is independent of the particular species, for all species considered in this work.

\section{Results and Discussion}
To verify the predicted detection efficiency experimentally, the number of {Ca}$^+$ ions in a given crystal is determined for several hundred Coulomb crystals of various sizes by comparing the experimental image immediately prior to ejection to a set of simulated crystal images (see Fig. \ref{fig:CCMD_sims}).  Following crystal ejection, the area under the relevant mass peak is calculated by integration of the (background-subtracted) TOF  trace.  Fig. \ref{fig:realidealtrappingvoltage}(a) illustrates the strong linear correlation between the number of ions in a Ca$^+$ Coulomb crystal and the integrated TOF peak recorded upon crystal ejection, for up to 200-ion crystals. The measurement error is governed by the uncertainty in the comparison between an experimental crystal and its MD simulation ($\pm5$ ions).    
The resolution of the mass spectrum, defined as $m/\Delta m$ where $\Delta m$ is the FWHM of the mass spectrum peak, is 90 for a typical 100-ion Ca$^+$ crystal and $>$~70 for all Ca$^+$ crystals containing up to 200 ions:  
 The slightly lower resolution for larger crystals is due to the greater spread of initial positions and velocities in those cases.
Crystals with more than 200 ions were not considered experimentally due to imaging limitations; the objective lens and camera restrict the field of view.  

\begin{figure}[!htb]
\centering
\vbox{
    \includegraphics{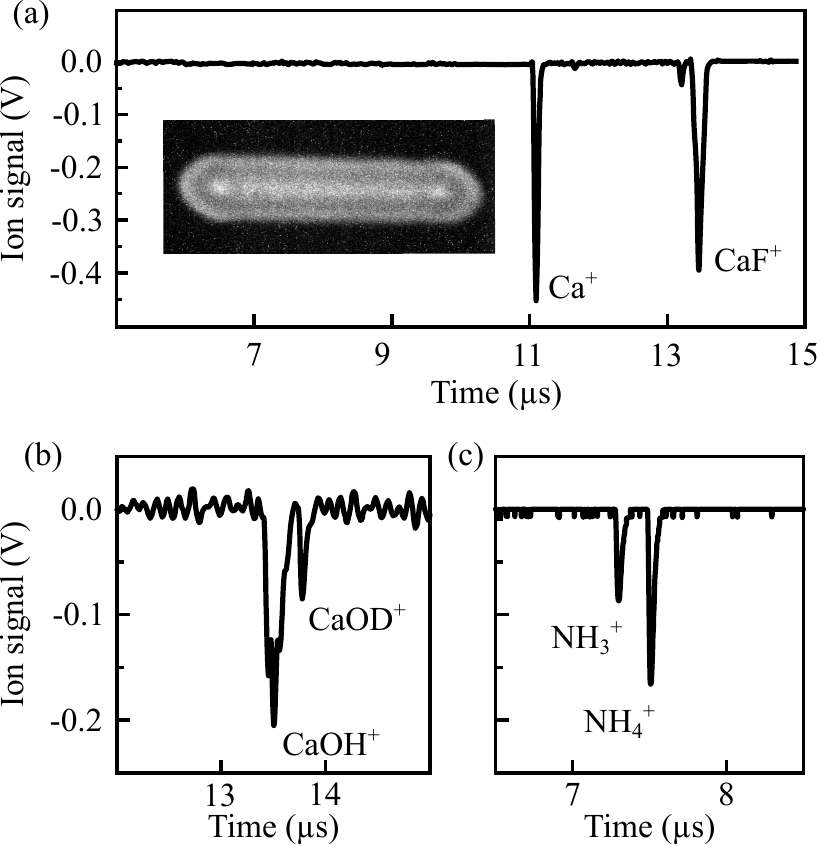}
}
\caption{(a) A bi-component Ca$^+$/CaF$^+$ Coulomb crystal and the TOF trace recorded following ejection of the crystal. Peaks arising from species with masses differing by 1 amu can be resolved, as demonstrated by the ejection of (b) Ca$^+$/CaOH$^+$/CaOD$^+$ and (c) NH$_3^+$/NH$_4^+$/Ca$^+$ multi-component crystals. 
 The two unlabelled small peaks in (a) are due to unidentified minor impurities.}
\label{fig:TOF_traces}
\end{figure}

To establish the applicability of this DIT MS technique to multi-component Coulomb crystals, mixed crystals of Ca$^+$ and CaF$^+$ are prepared by admitting a low pressure of  CH$_3$F gas to the ion trap chamber, leading to the reaction Ca$^+$ + CH$_3$F $\rightarrow$ CaF$^+$ + CH$_3$.  The resulting CaF$^+$ product ions sympathetically cool into the Ca$^+$ Coulomb crystal, forming a dark outer shell -- indicated by the flattening of the fluorescing Ca$^+$ core in the crystal images (see Figs. \ref{fig:CCMD_sims}(d) and \ref{fig:TOF_traces}(a)). Repeating simulations for variable numbers of laser-cooled ions and dark ions enables a determination of the number of each species in the crystal, assuming there is only one ``dark" component. 
The mass-to-charge ratio of CaF$^+$:Ca$^+$ is 1.475, well within the desired range for optimal sympathetic cooling.\cite{Willitsch_intrevPC,ARPC}  To our knowledge, this represents the first demonstration that multi-component Coulomb crystals can be formed in a DIT.

\begin{figure}[htb]
\centering
\vbox{
    \includegraphics{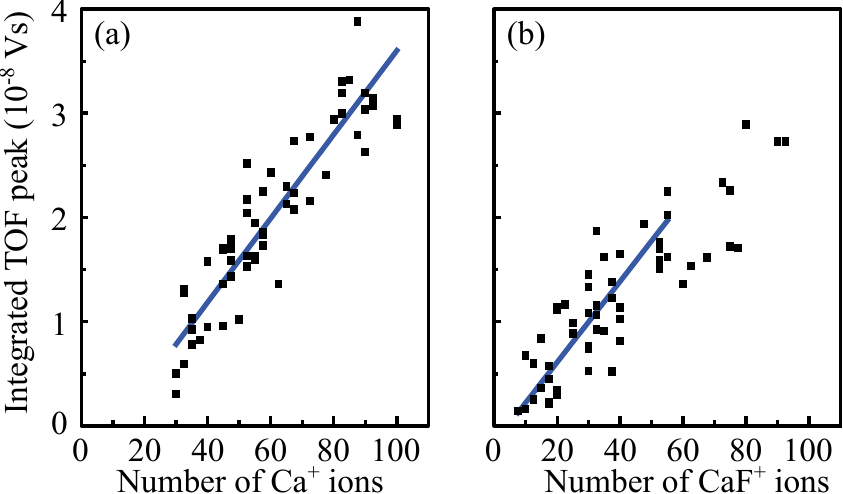}
}
\caption{Bi-component Ca$^+$/CaF$^+$ crystal ejection analysis. The integrated TOF peak for (a) Ca$^+$ and (b) CaF$^+$ ions is plotted against the number of ions of that species in each crystal,  ascertained from fitting the crystal images to simulations, for bi-component crystals of various sizes and compositions.}
\label{fig:Ca_CaF_linearfit}
\end{figure}

As Fig. \ref{fig:Ca_CaF_linearfit} illustrates, a linear relationship is observed between the integrated TOF mass peaks and ion numbers -- for both 
 the fluorescing
Ca$^+$ and 
the non-fluorescing 
CaF$^+$ ions.  The gradients of the lines-of-best-fit, $(4.0\pm0.2)\times$10$^{-10}$~Vs~ion$^{-1}$ for Ca$^+$ and $(3.9\pm0.4)\times$10$^{-10}$~Vs~ion$^{-1}$ for CaF$^+$ (for up to 60 CaF$^+$ ions), established using a linear least-squares fitting procedure with errors considered in both coordinates, are in quantitative agreement. 
These results are also in excellent agreement with 
the 
previous Ca$^+$-only 
 crystal 
results 
shown in figure \ref{fig:realidealtrappingvoltage}(a)), demonstrating the utility of the technique: equal detection efficiency is observed regardless of ion mass. There are too few crystals in the data set with $\ge$ 60 CaF$^+$ ions to extend the trend further, and the precision with which CaF$^+$ ion numbers can be estimated from simulations decreases at higher CaF$^+$:Ca$^+$ ratios.  Trajectory simulations indicate that the observed 
linear 
trend will hold for larger numbers of ions (up to 400-ion crystals), and for a diverse range of ionic species, provided the mass-to-charge ratios of co-trapped species are similar and there are a sufficient number of laser-cooled ions for efficient sympathetic cooling.

The mass resolution 
 of the technique
is verified by the ejection of crystals containing species differing in mass by 1~amu.  Tri-component Ca$^+$/CaOH$^+$/CaOD$^+$ crystals are formed in a two-step process.  H$_2$O vapor is admitted to the chamber through a leak valve, where reaction with Ca$^+$ (in the $3d$ $^2D$ excited states populated by the laser-cooling process)\cite{Okada} yields CaOH$^+$ molecular ions.  D$_2$O is then admitted through a second leak valve, forming CaOD$^+$ ions.  
As figure \ref{fig:TOF_traces}(b) illustrates, peaks arising from CaOH$^+$ and CaOD$^+$ are resolved in the spectrum, indicating a resolution better than 1 mass unit at 58 amu.

The applicability of this DIT MS technique for species lighter than Ca$^+$ is confirmed with tri-component NH$_3^+$/NH$_4^+$/Ca$^+$ Coulomb crystals.  NH$_3$ is resonantly ionized at 312.5~nm, with some 
 of the sympathetically cooled
NH$_3^+$ ions subsequently reacting with neutral NH$_3$ to form NH$_4^+$, yielding tri-component  crystals.  Following the same imaging and ejection process detailed above, the TOF spectra again exhibit well-resolved peaks, as shown in figure \ref{fig:TOF_traces}(c).  In this system, both the dark species are lighter than Ca$^+$.  Accordingly, conventional comparisons of the experimental crystals with simulated images do not reveal the relative numbers of NH$_3^+$ and NH$_4^+$ ions, as both species are located in the dark inner core. The same applies to multi-component Ca$^+$/CaOH$^+$/CaOD$^+$ crystals, where the CaOH$^+$ and CaOD$^+$ species are 
both heavier than Ca$^+$ and therefore are both located in the dark outer shell.  DIT MS enables us to acertain the relative numbers of dark ions which are otherwise indistinguishable in the experimental and simulated images.

\section{Conclusions}
In summary, we have established DIT MS to be a robust technique for the quantitative characterization of multi-component Coulomb crystals produced, in this work, by chemical reactions of the trapped ions.  The ejection process is efficient, with a linear correlation 
between the number of ions ejected and the corresponding integrated TOF peak, and a detection efficiency that is independent of the crystal size or the ionic species involved. 
The DIT MS technique is applicable to a diverse range of multi-component Coulomb crystals
and hence to a diverse range of chemical reactions -- not only reactions of laser cooled ions but also of sympathetically cooled ions.
  While the ejection process is destructive, repeating the process with reloaded crystals at a range of reaction times allows the determination of accurate reaction rates and branching ratios in complicated ion-molecule reactions.  With the ability to unambiguously detect both the masses and relative numbers of trapped ions, DIT MS has the potential to be a powerful detection technique in the study of sympathetically-cooled molecular ion reactions -- utilized as a stand-alone method or in combination with existing techniques such as MD simulations.

\begin{acknowledgments}
TPS acknowledges the financial support of the EPSRC under grants EP/G00224X/1 and EP/1029109. ND is grateful for support from the Felix Scholarships Foundation, LLP from the Clarendon Fund, CJR from the Ramsay Memorial Trust, and BRH from the Royal Commission for the Exhibition of 1851 and the EU Marie Curie Career Integration Grant scheme (PCIG13-GA-2013-618156). 
\end{acknowledgments}

%

\end{document}